\documentclass[a4paper,12pt]{article}
\usepackage{amsmath,amsthm}
\usepackage[letterpaper,margin=1in]{geometry}
\usepackage{blindtext}
\usepackage[symbol]{footmisc}
\numberwithin{equation}{section}

\usepackage[colorlinks,citecolor=blue,urlcolor=red,bookmarks=false,hypertexnames=true]{hyperref}
\DeclareRobustCommand{\rchi}{{\mathpalette\irchi\relax}}
\newcommand{\irchi}[2]{\raisebox{\depth}{$#1\chi$}} 
\begin{document}
\centerline{}\centerline{\Large{\bf A Study On Some Geometric and
Physical Properties of }} \centerline{}\centerline{\Large{\bf
Hyper-Generalised Quasi-Einstein Spacetime }}
\centerline{}\centerline{$~$Kaushik
Chattopadhyay\textsuperscript{1}\footnote{This is the
corresponding author.}, Arindam Bhattacharyya\textsuperscript{2}
and Dipankar Debnath\textsuperscript{3}}
\centerline{\textsuperscript{1}Department of
Mathematics,}\centerline{Jadavpur University, Kolkata-700032,
India} \centerline{E-mail:kausikchatterji@gmail.com}
\centerline{\textsuperscript{2}Department of
Mathematics,}\centerline{Jadavpur University, Kolkata-700032,
India} \centerline{E-mail:bhattachar1968@yahoo.co.in}
\centerline{\textsuperscript{3}Department of
Mathematics,}\centerline{Bamanpukur High School(H.S), Nabadwip,
India} \centerline{E-mail:dipankardebnath123@hotmail.com}
\newtheorem{Theorem}{Theorem}[section]
\begin{abstract}{\em In the present paper we discuss about a set of geometric and physical properties of hyper-generalised quasi-Einstein spacetime. At the beginning we discuss about pseudosymmetry over a hyper-generalised quasi-Einstein spacetime. Here we discuss about $W_{2}$-Ricci pseudosymmetry, $Z$-Ricci pseudosymmetry,  Ricci pseudosymmetry and projective pseudosymmetry over a hyper-generalised quasi-Einstein spacetime. Later on we take over Ricci symmetric hyper-generalised quasi-Einstein spacetime and derive a set of important geometric and physical theorems over it. Moving further we consider some physical applications of the hyper-generalised quasi-Einstein spacetime. Lastly we prove the existence of a hyper-generalised quasi-Einstein spacetime by constructing a non-trivial example.}
\end{abstract}
\textbf{M.S.C.2010:} ~~53C15, 53C25, 53C35.

\section{Introduction}

$~~~~$The General theory of Relativity is unarguably the most beautiful theory the World of Physics has ever produced. It is the most powerful result of the human intellect. This is an extremely important theory to study the nature of this universe, cosmology and gravity. Three most important things that Modern Scientists/ Mathematical Physicists can learn from special to general relativity are as follows:\\
$~~~~~$(i) The laws of Physics should be the same in every inertial reference frame, i. e., the abandonment of the privileged states of inertial frame of reference.\\
$~~~~~$(ii) The acceptance of the dynamical role of the metric $g$, i.e., the study of non-linear behaviour of nature.\\
and\\
$~~~~~$(iii) The spacetime has to be considered as a class of semi-Riemannian geometry.\\
The semi-Riemannian geometry has become more and more relevant and significant in dealing with the nature of this universe with every passing day.\\\\
The theory of general relativity is mainly studied on a semi-Riemannian manifold which sometimes is not an Einstein spacetime. Thus it was always necessary to expand the concept of Einstein manifold to quasi-Einstein, then generalised quasi-Einstein, mixed generalised quasi-Einstein and lastly to hyper-generalised quasi-Einstein manifold. We demonstrate the introduction to this procedure as follows:\\\\
$~~~$An Einstein manifold is a Riemannian or pseudo-Riemannian manifold whose Ricci tensor S of type $(0, 2)$ is non-zero and proportional to the metric tensor. Einstein manifolds form a natural subclass of various classes of Riemannian or semi-Riemannian manifolds by a curvature condition imposed on their Ricci tensor $\cite{jmh1}$. Also in Riemannian geometry as well as in general relativity theory, the Einstein manifold plays a very important role.\\\\
$~~~~$M. C. Chaki and R. K. Maity had given the notion of quasi
Einstein manifold $\cite{jmh2}$ in $2000$. A non flat
$n$-dimensional Riemannian manifold $(M^{n},g)$, $n(>2)$ is said
to be a quasi-Einstein manifold if its nonzero Ricci tensor $S$ of
type $(0,2)$ satisfies the following condition
\begin{equation} \label{e1.1}
S(X,Y)=\alpha g(X,Y)+\beta A(X)A(Y),
\end{equation}
where for all vector fields $X$,
\begin{equation} \label{e1.2}
g(X,\xi_{1})= A(X),~g(\xi_{1},\xi_{1})= 1.
\end{equation}
That is, $A$ being the associated $1$-form, $\xi_{1}$ is generally known as the generator of the manifold. $\alpha$ and $\beta$ are associated nonzero scalar functions. This manifold is denoted by  $(QE)_n$. Clearly, for $\beta= 0$, this manifold reduces an Einstein manifold. We can note that Robertson-Walker spacetimes are quasi-Einstein spacetimes. In the recent papers $\cite{jmh4}$, $\cite{jmh3}$, the application of quasi-Einstein spacetime and generalised quasi-Einstein spacetime in general relativity have been studied. Many more works have been done in the spacetime of  general relativity $\cite{jmh31}$, $\cite{jmh41}$, $\cite{jmh22}$, $\cite{jmh7}$, $\cite{jmh8}$, $\cite{jmh38}$, $\cite{jmh27}$, $\cite{jmh28}$, $\cite{jmh5}$.\\\\
$~~~~$Then M. C. Chaki initiated the notion of generalized
quasi-Einstein manifold $\cite{jmh9}$ in 2001. A Riemannian
manifold of dimension $n(>2)$ is said to be generalized quasi
Einstein manifold if its Ricci tensor $S$ of type $(0,2)$ is not
identically zero and satisfies the following condition
\begin{equation}  \label{e1.3}
S(X,Y)=\alpha g(X,Y)+\beta A(X)A(Y)+\gamma[A(X)B(Y)+A(Y)B(X)],
\end{equation}
where $\alpha$, $\beta$ and $\gamma$ are real valued, nonzero
scalar functions on $(M^n,g)$, $A$ and $B$ are called two non zero
$1$-forms such that
\begin{equation} \label{e1.4}
g(X,\xi_{1})=A(X), g(X,\xi_{2})=B(X),
g(\xi_{1},\xi_{2})=0,g(\xi_{1},\xi_{1})=1, g(\xi_{2},\xi_{2})=1.
\end{equation}
Here $\xi_{1}$ and $\xi_{2}$ are two unit vector fields which are orthogonal to each other. $\alpha$, $\beta$ and $\gamma$ are called associated scalars, $A$ and $B$ are called associated $1$-forms. $\xi_{1}$ and $\xi_{2}$ are two generators of the manifold. This manifold is denoted by $(GQE)_{n}$. Clearly, for $\gamma=0$, then it takes the form of a quasi-Einstein manifold and for $\beta=\gamma=0$, it takes the form of an Einstein manifold.\\\\
$~~~~$The notion of hyper-generalized quasi-Einstein manifold has
been introduced by A. A. Shaikh, C. \"{O}zg\"{u}r and A. Patra
$\cite{jmh21}$ in 2011. According to them, a Riemannian manifold
$(M^{n},g)$, $(n>2)$ is said to be a hyper-generalized
quasi-Einstein manifold if its Ricci tensor S of type $(0, 2)$ is
non-zero and satisfies the following condition
\begin{eqnarray} \label{e1.5}
&&S(X,Y)=\alpha g(X,Y)+\beta A(X)A(Y)+\gamma[A(X)B(Y)+A(Y)B(X)]\nonumber\\
&&~~~~~~~~~~~+\delta[A(X)D(Y)+A(Y)D(X)],
\end{eqnarray}
for all $X, Y, Z \in \rchi (M)$. Here $\alpha$, $\beta$, $\gamma$,
$\delta$ are non-zero scalar functions on $(M^{n}, g)$. $A$, $B$,
$D$ are non-zero $1$-forms such that
\begin{equation} \label{e1.6}
g(X, \xi_{1})=A(X),~g(X, \xi_{2})=B(X),~g(X, \xi_{3})=D(X),
\end{equation}
where $\xi_{1}$, $\xi_{2}$, $\xi_{3}$ are mutually orthogonal unit
vector fields. i. e.,
\begin{equation} \label{e1.7}
g(\xi_{1},\xi_{2})=g(\xi_{2},\xi_{3})=g(\xi_{1},\xi_{3})=0;~g(\xi_{1},\xi_{1})=g(\xi_{2},\xi_{2})=g(\xi_{3},\xi_{3})=1.
\end{equation}
 An $n$-dimensional hyper-generalized quasi-Einstein manifold is generally denoted as $(HGQE)_{n}$. Shaikh, \"{O}zg\"{u}r and Patra in $\cite{jmh21}$ studied on hyper-generalized quasi-Einstein manifolds with some geometric properties of it. G\"{u}ler and Demirba\v{g} $\cite{jmh14}$ dealt with some Ricci conditions on hyper-generalized quasi-Einstein manifolds, D. Debnath $\cite{jmh37}$ proved few theorems about the properties of the hyper-generalized quasi-Einstein manifolds.\\\\
$~~~~$The concept of perfect fluid spacetime arose while
discussing the structure of this universe. Perfect fluids are
often used in the general relativity to model the idealised
distribution of matter, such as the interior of a star or
isotropic pressure. In general relativity the matter content of
the spacetime is described by the energy-momentum tensor. The
matter content is assumed to be a fluid having density and
pressure and possessing dynamical and kinematical quantities like
velocity, acceleration, vorticity, shear and expansion. The
energy-momentum tensor $T$ of a perfect fluid spacetime is given
by the following equation $\cite{jmh12}, \cite{jmh13}$
\begin{equation} \label{e1.8}
T(X,Y)= (\sigma+p)A(X)A(Y)+ pg(X,Y).
\end{equation}
Here $g(X,\xi_{1})=A(X), A(\xi_{1})= -1$, for any $X,Y$. $p$ and $\sigma$ are called the isotropic pressure and the energy density respectively. $\xi_{1}$ being the unit timelike velocity vector field.\\\\
The Einstein field equation $\cite{jmh30}$ is given by
\begin{equation} \label{e1.9}
S(X,Y)-\frac{r}{2}g(X,Y)+ \lambda g(X,Y)= kT(X,Y);~\forall X,Y \in
TM,
\end{equation}
here $r$ being the scalar curvature, $S$ being the Ricci tensor of type $(0,2)$. $k$ and $\lambda$ are the gravitational constant and cosmological constant respectively. From Einstein's field equation it follows that energy momentum tensor is a symmetric $(0,2)$ type tensor of divergence zero.\\\\
In the year of 2012, Mantica and Molinari$\cite{jmh34}$ defined a
new generalized symmetric $(0,2)$ tensor called Z tensor.
According to them it is given as,
\begin{equation}\label{e1.10}
Z(X,Y)=S(X,Y)+\phi g(X,Y),
\end{equation}
where $\phi$ is an arbitrary scalar function. A set of properties of $Z$ tensor have been studied in the papers $\cite{jmh34}$ and $\cite{jmh35}$.\\\\
Like the $Z$ curvature tensor projective curvature tensor also
plays a very significant role in studying different properties of
semi-Riemannian geometry. Let $M^{n}(n \ge 3)$ be a
semi-Riemannian manifold. The projective curvature
tensor$\cite{jmh32}$ is defined by,
\begin{equation}\label{e1.11}
P(X,Y)Z=R(X,Y)Z-\frac{1}{n-1}\{S(Y,Z)X-S(X,Z)Y\}.
\end{equation}
$~~~~$Hyper-generalized quasi-Einstein manifolds is considered as the base space of general relativistic viscous fluid spacetime, which inspired us to take a look on some geometric properties of the $(HGQE)_{4}$ spacetime under certain conditions which we study on sections \ref{s2}, \ref{s3}, \ref{s4} and \ref{s5}. Then we discuss about Ricci symmetric hyper-generalized quasi-Einstein spacetimes in section \ref{s6}. In section \ref{s7} we derive a result about the energy-momentum tensor on hyper-generalized quasi-Einstein spacetime of constant curvature with cyclic parallel Ricci tensor. Also the spacetime has wide applications in general relativistic viscous fluid spacetime admitting heat flux and stress, which motivated us to discuss about some physical applications of an $(HGQE)_{4}$ spacetime in section \ref{s8}. Finally in section \ref{s9} we construct a non-trivial example of an $(HGQE)_{4}$ spacetime to prove the existence of such spacetime.\\

\section{$W_{2}$-Ricci pseudosymmetric $(HGQE)_{4}$ spacetime} \label{s2}

$~~~~$The $W_{2}$-curvature tensor was introduced by G. P.
Pokhariyal and R. S. Mishra $\cite{jmh10}$ in $1970$ and they
studied some properties of it. A $W_{2}$-curvature tensor on a
manifold  $(M^{n},g)$, $n(>3)$ is defined by
\begin{equation} \label{e2.1}
W_{2}(X,Y)Z= R(X,Y)Z-\frac{1}{n-1}[g(Y,Z)QX-g(X,Z)QY].
\end{equation}
Here $r$ being the curvature tensor and $Q$ is the Ricci operator defined by $g(QX,Y)= S(X,Y),\forall~X,Y$.\\\\
For an $(HGQE)_{4}$ quasi-Einstein spacetime the $(0,4)$ $W_{2}$
curvature tensor takes the following form,
\begin{equation} \label{e2.2}
W_{2}(X,Y,Z,W)=R(X,Y,Z,W)-\frac{1}{3}[g(Y,Z)S(X,W)-g(X,Z)S(Y,W)].
\end{equation}
Firstly, we take a hyper generalized quasi-Einstein spacetime
satisfying the condition $W_{2}. S = F_{S} Q(g,S)$. Here $F_{S}$
being a certain function on the set $U_{S} = \{x\in M :S\neq
\frac{r}{n}g$ at $x\}$ and $Q(g,S)$ being the Tachibana tensor
working on the metric tensor and the Ricci tensor. This spacetime
is called $ W_{2}$-Ricci pseudosymmetric $ (HGQE)_{4}$. Now for
all $ X,Y,Z \in \rchi(M^4)$;
\begin{eqnarray} \label{e2.3}
& & S(W_{2}(X,Y)Z,W)+ S(Z,W_{2}(X,Y)W)\nonumber\\
&&= F_{S}[g(Y,Z)S(X,W)-g(X,Z)S(Y,W)\nonumber\\
&&+ g(Y,W)S(Z,X)- g(X,W)S(Y,Z)].
\end{eqnarray}
From the equation (\ref{e1.5}) from the equation (\ref{e2.3}) we
get,
\begin{eqnarray} \label{e2.4}
&&\alpha g(W_{2}(X,Y)Z,W)+\beta A(W_{2}(X,Y)Z)A(W)\nonumber\\
&&+\gamma[A(W_{2}(X,Y)Z)B(W)+B(W_{2}(X,Y)Z)A(W)] \nonumber\\
&&+\delta[A(W_{2}(X,Y)Z)D(W)+D(W_{2}(X,Y)Z)A(W)]\nonumber\\
&&+\alpha g(W_{2}(X,Y)W,Z)+\beta A(W_{2}(X,Y)W)A(Z)\nonumber\\
&&+\gamma [A(W_{2}(X,Y)W)B(Z)+B(W_{2}(X,Y)W)A(Z)] \nonumber\\
&&+\delta [A(W_{2}(X,Y)W)D(Z)+D(W_{2}(X,Y)W)A(Z)] \nonumber\\
&&=F_{S}[g(Y,Z)S(X,W)-g(X,Z)S(Y,W) \nonumber\\
&&~~+g(Y,W)S(Z,X)-g(X,W)S(Y,Z)].
\end{eqnarray}
Contracting the equation over $X$ and $W$ and putting $Z=\xi_{1}$
we get,
\begin{eqnarray}\label{e2.5}
&~&\alpha[\frac{4}{3}\{(\alpha-\beta)A(Y)-\gamma B(Y)-\delta D(Y)\}-\frac{r}{3}A(Y)]-\frac{1}{3}[-\gamma B(Y)-\delta D(Y)]\nonumber\\
&&+\frac{\gamma^{2}}{3}A(Y)+\gamma R(\xi_{1},Y,\xi_{1},\xi_{2})-\frac{\alpha \gamma}{3}B(Y)+\frac{\delta^{2}}{3}A(Y)\nonumber\\
&&+\delta R(\xi_{1},Y,\xi_{1},\xi_{3})-\frac{\delta \alpha}{3} D(Y)\nonumber\\
&&=F_{S}[rA(Y)-4\{(\alpha-\beta)A(Y)-\gamma B(Y)-\delta D(Y)\}].
\end{eqnarray}
Putting $X=\xi_{1}, Z=\xi_{2}, W=\xi_{3}$ in the equation
(\ref{e2.4}) we get,
\begin{eqnarray}\label{e2.6}
&~&\delta R(\xi_{1},Y,\xi_{2},\xi_{1})+\gamma R(\xi_{1},Y,\xi_{3},\xi_{1})-\frac{\alpha}{3}B(Y)(-\delta)\nonumber\\
&&-\frac{\alpha}{3}D(Y)(-\gamma)+\frac{\delta}{3}B(Y)(\alpha-\beta)+\frac{\gamma}{3}D(Y)(\alpha-\beta)\nonumber\\
&&=F_{S}[B(Y)(-\delta)+D(Y)(-\gamma)].
\end{eqnarray}
combining the equations (\ref{e2.5}) and (\ref{e2.6}) we get,
\begin{eqnarray}\label{e2.7}
&~&\alpha[\frac{4}{3}\{(\alpha-\beta)A(Y)-\gamma B(Y)-\delta D(Y)\}-\frac{r}{3}A(Y)]-\frac{1}{3}[-\gamma B(Y)-\delta D(Y)]\nonumber\\
&&+\frac{\gamma^{2}}{3}A(Y)-\frac{\alpha \gamma}{3}B(Y)+\frac{\delta^{2}}{3}A(Y)\nonumber\\
&&-\frac{\delta \alpha}{3} D(Y)-F_{S}[B(Y)(-\delta)+D(Y)(-\gamma)]-\frac{\alpha}{3}B(Y)(-\delta)\nonumber\\
&&-\frac{\alpha}{3}D(Y)(-\gamma)+\frac{\delta}{3}B(Y)(\alpha-\beta)+\frac{\gamma}{3}D(Y)(\alpha-\beta)\nonumber\\
&&=F_{S}[rA(Y)-4\{(\alpha-\beta)A(Y)-\gamma B(Y)-\delta D(Y)\}].
\end{eqnarray}
setting $\gamma=\delta$ yields,
\begin{eqnarray}\label{e2.7}
&~&\alpha[\frac{4}{3}\{(\alpha-\beta)A(Y)-\gamma B(Y)-\gamma D(Y)\}-\frac{r}{3}A(Y)]-\frac{1}{3}[-\gamma B(Y)-\gamma D(Y)]\nonumber\\
&&+\frac{\gamma^{2}}{3}A(Y)-\frac{\alpha \gamma}{3}B(Y)+\frac{\gamma^{2}}{3}A(Y)\nonumber\\
&&-\frac{\gamma \alpha}{3} D(Y)-F_{S}[B(Y)(-\gamma)+D(Y)(-\gamma)]-\frac{\alpha}{3}B(Y)(-\gamma)\nonumber\\
&&-\frac{\alpha}{3}D(Y)(-\gamma)+\frac{\gamma}{3}B(Y)(\alpha-\beta)+\frac{\gamma}{3}D(Y)(\alpha-\beta)\nonumber\\
&&=F_{S}[rA(Y)-4\{(\alpha-\beta)A(Y)-\gamma B(Y)-\gamma D(Y)\}].
\end{eqnarray}
Putting $Y=\xi_{1}$ in the above equation we get,
\begin{equation}\label{e2.8}
-3\beta F_{S}=\alpha \beta -\frac{2\gamma^{2}}{3}.
\end{equation}
Again putting $Y=\xi_{2}$ in the equation (\ref{e2.7}) we get,
\begin{equation}\label{e2.9}
\gamma(\alpha+3F_{S})=0,
\end{equation}
which gives either $\gamma=0$ or $F_{S}=\frac{-\alpha}{3}$. Now, if $\gamma=0$ then since $\gamma=\delta$ thus $\gamma=\delta=0$. Hence the spacetime becomes a quasi-Einstein spacetime. On the other hand if $F_{S}=\frac{-\alpha}{3}$ then from the equation (\ref{e2.8}) we get $\gamma=0$. Again since $\gamma=\delta$ thus $\gamma=\delta=0$. Thus in both the cases the manifold is reduced to a quasi-Einstein spacetime. Hence we conclude the following theorem as:\\\\
\textbf{Theorem 2.1:} {\em A $W_{2}$-Ricci pseudosymmetric
$(HGQE)_{4}$ spacetime is a $(QE)_{4}$ spacetime if
$\gamma=\delta$.}

\section{Z-Ricci pseudosymmetric $(HGQE)_{4}$ spacetime} \label{s3}
A semi-Riemannian manifold $(M^{n},g),n \ge 3$, is called Z-Ricci
pseudosymmetric iff the following relation holds:
\begin{equation}\label{3.1}
Z\cdot K=F_{K}P(g,K),
\end{equation}
on the set $U_{K}=\{x \in M: P(g, K) \neq 0~at~x\}$, where $K$ is
the Ricci operator defined by $S(X,Y)=g(KX,Y)$ and $F_{K}$ is a
smooth function on $U_{K}$. The operator $P$ is defined by the
following way:
\begin{equation}\label{e3.2}
P(g,K)(W;X,Y)=K((X\wedge_{g} Y)W)
\end{equation}
for all vector fields $X,Y,W$.\\\\
Now if the spacetime is a Z-Ricci pseudosymmetric then from the
equation (\ref{3.1}) we get,
\begin{equation}\label{e3.3}
(Z(X,Y)\cdot K)W=F_{K}P(g,K)(W;X,Y),
\end{equation}
which takes the following form,
\begin{eqnarray} \label{e3.4}
&&Z(Y,KW)X-Z(X,KW)Y-Z(Y,W)KX-Z(X,W)KY \nonumber\\
&&=F_{K}\{g(Y,W)KX-g(X,W)KY\}.
\end{eqnarray}
With the help of the equation $(\ref{e1.5})$ we demonstrate the
Ricci operator by the following equation:
\begin{equation}\label{e3.5}
KX=\alpha X+\beta
A(X)\xi_{1}+\gamma[A(X)\xi_{2}+B(X)\xi_{1}]+\delta[A(X)\xi_{3}+D(X)\xi_{1}].
\end{equation}
Applying the equation (\ref{e3.5}) in equation (\ref{e3.4}) we
get,
\begin{eqnarray}\label{e3.6}
&~&Z(Y,\alpha W)X+\beta A(W)Z(Y,\xi_{1})X+\gamma A(W)Z(Y,\xi_{2})X\nonumber\\
&&+\gamma B(W)Z(Y,\xi_{1})X+\delta A(W)Z(Y,\xi_{3})X+\delta D(W)Z(Y,\xi_{1})X\nonumber\\
&&-[Z(X,\alpha W)Y+\beta A(W)Z(X,\xi_{1})Y+\gamma A(W)Z(X,\xi_{2})Y\nonumber\\
&&+\gamma B(W)Z(X,\xi_{1})Y+\delta A(W)Z(X,\xi_{3})Y+\delta D(W)Z(X,\xi_{1})Y]\nonumber\\
&&=\{F_{K}g(Y,W)+Z(Y,W)\}KX-\{F_{K}g(X,W)-Z(X,W)\}KY.
\end{eqnarray}
With the help of the equation (\ref{e1.10}) this further implies,
\begin{eqnarray}\label{e3.7}
&~&\alpha Z(Y,W)X+\beta A(W)\{(\alpha-\beta+\phi)A(Y)-\gamma B(Y)-\delta D(Y)\}X\nonumber\\
&&+\gamma A(W)\{(\alpha+\phi)B(Y)+\gamma A(Y)\}X\nonumber\\
&&+\gamma B(W)\{(\alpha-\beta+\phi)A(Y)-\gamma B(Y)-\delta D(Y)\}X\nonumber\\
&&+\delta A(W)\{(\alpha+\phi)D(Y)+\gamma A(Y)\}X\nonumber\\
&&+\delta D(W)\{(\alpha-\beta+\phi)A(Y)-\gamma B(Y)-\delta D(Y)\}X\nonumber\\
&&-\alpha Z(X,W)Y-\beta A(W)\{(\alpha-\beta+\phi)A(X)-\gamma B(X)-\delta D(X)\}Y\nonumber\\
&&-\gamma A(W)\{(\alpha+\phi)B(X)+\gamma A(X)\}Y\nonumber\\
&&-\gamma B(W)\{(\alpha-\beta+\phi)A(X)-\gamma B(X)-\delta D(X)\}Y\nonumber\\
&&-\delta A(W)\{(\alpha+\phi)D(X)+\gamma A(X)\}Y\nonumber\\
&&-\delta D(W)\{(\alpha-\beta+\phi)A(X)-\gamma B(X)-\delta D(X)\}Y\nonumber\\
&&=\{F_{K}g(Y,W)+S(Y,W)+\phi g(Y,W)\}KX\nonumber\\
&&-\{F_{K}g(X,W)-S(X,W)-\phi g(X,W)\}KY.
\end{eqnarray}
Putting $X=\xi_{1}, Y=\xi_{2}$ equation (\ref{e3.7}) yields,
\begin{eqnarray}\label{e3.8}
&~&\alpha \{(\alpha+\phi)B(W)+\gamma A(W)\}\xi_{1}+\beta A(W)\{-\gamma \}\xi_{1}+\gamma A(W) \{\alpha+\phi\}\xi_{1}\nonumber\\
&&+\gamma B(W)\{-\gamma\}\xi_{1}+\delta D(W)\{-\gamma\}\xi_{1}-\alpha \{(\alpha-\beta+\phi)A(W)-\gamma B(W)-\delta D(W)\}\xi_{2}\nonumber\\
&&+\beta A(W)\{\alpha-\beta+\phi)\}\xi_{2}+\gamma^{2}A(W)\xi_{2}\nonumber\\
&&+\gamma B(W)\{\alpha-\beta+\phi)\}\xi_{2}+\delta^{2}A(W)\xi{2}+\delta D(W)\{\alpha-\beta+\phi)\}\xi_{2}\nonumber\\
&&=\{F_{K}B(W)+\phi B(W)+\alpha B(W)+\gamma A(W)\}K\xi_{1}\nonumber\\
&&-\{F_{K}A(W)-\phi A(W)-(\alpha-\beta)A(W)+\gamma B(W)+\delta
D(W)\}K\xi_{2}.
\end{eqnarray}
Taking inner product with $\xi_{1}$ to both the sides of the
equation (\ref{e3.8}) we get,
\begin{eqnarray}\label{e3.9}
A(W)\{\alpha \gamma+\phi
\gamma\}+B(W)\{-\gamma^{2}-F_{K}(\alpha-\beta)+\phi \beta +\alpha
\beta\}-D(W)\{\delta \gamma\}=0.
\end{eqnarray}
Putting $W=\xi_{3}$ in equation (\ref{e3.9}) we get,
\begin{equation}
-\gamma \delta=0.
\end{equation}
That means at least one of $\gamma$ or $\delta$ must be zero. Which means the manifold is reduced to a generalized quasi-Einstein spacetime. This allows us to derive the next theorem as:\\\\
\textbf{Theorem 3.1:} {\em A Z-Ricci pseudosymmetric $(HGQE)_{4}$
spacetime is a $(GQE)_{4}$ spacetime.}

\section{Ricci pseudosymmetric $(HGQE)_{4}$ spacetime} \label{s4}
A semi-Riemannian manifold $M^{n}(n \ge 3)$ is called
Ricci-pseudosymmetric if the following relation
\begin{equation}\label{e4.1}
(R(X,Y) \cdot S)(Z,W)=F_{S}Q(g,S)
\end{equation}
holds on $U_{S}=\{x \in M: S \neq \frac{r}{n}~at~x\}$ and $L_{S}$
is a function on $U_{S}$. From the equation (\ref{e4.1}) we get,
\begin{eqnarray}\label{e4.2}
&&~~S(R(X,Y)Z,W)+S(Z,R(X,Y)W)\nonumber\\
&&=F_{S}[g(Y,Z)S(X,W)-g(X,Z)S(Y,W)\nonumber\\
&&+g(Y,W)S(Z,X)-g(X,W)S(Y,Z)].
\end{eqnarray}
Using the equation (\ref{e1.5}) we demonstrate the equation
(\ref{e4.2}) as follows:
\begin{eqnarray}\label{e4.3}
&&~~\beta[A(R(X,Y)Z)A(W)+A(Z)A(R(X,Y)W)]\nonumber\\
&&+\gamma[A(R(X,Y)Z)B(W)+A(W)B(R(X,Y)Z)+A(Z)B(R(X,Y)W)\nonumber\\
&&+A(R(X,Y)W)B(Z)]+\delta[A(R(X,Y)Z)D(W)+A(W)D(R(X,Y)Z)\nonumber\\
&&+A(Z)D(R(X,Y)W)+A(R(X,Y)W)D(Z)]\nonumber\\
&&=F_{S}[\beta\{g(Y,Z)A(X)A(W)-g(X,Z)A(Y)A(W)+g(Y,Z)A(Z)A(X)\nonumber\\
&&-g(X,W)A(Y)A(Z)\}+\gamma\{g(Y,Z)[A(X)B(W)+A(W)B(X)]\nonumber\\
&&-g(X,Z)[A(Y)B(W)+A(W)B(Y)]+g(Y,W)[A(X)B(Z)+A(Z)B(X)]\nonumber\\
&&-g(X,W)[A(Y)B(Z)+A(Z)B(Y)]\}+\delta\{g(Y,Z)[A(X)D(W)+A(W)D(X)]\nonumber\\
&&-g(X,Z)[A(Y)D(W)+A(W)D(Y)]+g(Y,Z)[A(X)D(Z)+A(Z)D(X)]\nonumber\\
&&-g(X,W)[A(Y)D(Z)+A(Z)D(Y)]\}].
\end{eqnarray}
Contracting equation (\ref{e4.3}) over X and W we get,
\begin{eqnarray}\label{e4.4}
&&~~\beta[A(R(\xi_{1},Y)Z)-A(Z)S(Y,\xi_{1})]+\gamma[A(R(\xi_{2},Y)Z)\nonumber\\
&&+B(R(\xi_{1},Y)Z)-A(Z)S(Y,\xi_{2})-B(Z)S(Y,\xi_{1})]\nonumber\\
&&+\delta[A(R(X,Y)Z)D(W)+D(R(\xi_{1},Y)Z)\nonumber\\
&&-A(Z)S(Y,\xi_{3})-S(Y,\xi_{1})D(Z)]\nonumber\\
&&=F_{S}\{\beta[-g(Y,Z)-4A(Y)A(Z)]-4\gamma[A(Y)B(Z)\nonumber\\
&&A(Z)B(Y)]-4\delta[A(Y)D(Z)+A(Z)D(Y)]\}.
\end{eqnarray}
Setting $Z=\xi_{1}$ in equation ($\ref{e4.4}$) we get,
\begin{eqnarray}\label{e4.5}
&&~~\beta(Y,\xi_{1})+\gamma[R(\xi_{1},Y,\xi_{1},\xi_{2})+S(Y,\xi_{2})]\nonumber\\
&&+\delta[R(\xi_{1},Y,\xi_{1},\xi_{3})+S(Y,\xi_{3})]\nonumber\\
&&=F_{S}[3\beta A(Y)+4\gamma B(Y)+4\delta D(Y)].
\end{eqnarray}
Now, putting $Z=\xi_{2}, W=\xi_{3}$ in the equation (\ref{e4.3})
we have,
\begin{eqnarray}\label{e4.6}
&&-\gamma R(\xi_{1},Y,\xi_{1},\xi_{3})-\delta R(\xi_{1},Y,\xi_{1},\xi_{2})\nonumber\\
&&=F_{S}\{\gamma[D(Y)A(X)-D(X)A(Y)]+\delta[A(X)B(Y)-A(Y)B(X)]\}.
\end{eqnarray}
If $\gamma=\delta$ then from the equations (\ref{e4.5}) and
(\ref{e4.6}) we conclude,
\begin{eqnarray}\label{e4.7}
&&[\beta S(Y,\xi_{1})+\gamma S(Y,\xi_{2})+\gamma S(Y,\xi_{3})]\nonumber\\
&&-F_{S}[3\beta A(Y)+4\gamma B(Y)+4\gamma D(Y)]\nonumber\\
&&=F_{S}\{\gamma[D(Y)A(X)-D(X)A(Y)]+\gamma[A(X)B(Y)-A(Y)B(X)]\}.
\end{eqnarray}
Putting $X=\xi_{1}$ we get,
\begin{eqnarray}\label{e4.8}
&&A(Y)[\alpha \beta-\beta^{2}+2\gamma^{2}]+B(Y)[-\beta \gamma+\gamma \alpha]+D(Y)[-\beta \gamma+\gamma \alpha]\nonumber\\
&&-F_{S}[3\beta A(Y)+4\gamma B(Y)+4\gamma
D(Y)]=-F_{S}\gamma[D(Y)+B(Y)].
\end{eqnarray}
Putting $Y=\xi_{2}$ in equation (\ref{e4.8}) we get,
\begin{equation}\label{e4.9}
\gamma[(\alpha-\beta)-3F_{S}]=0.
\end{equation}
Again putting $Y=\xi_{1}$ in equation (\ref{e4.8}) we get,
\begin{equation}\label{e4.10}
F_{S}=\frac{\alpha \beta-\beta^{2}+2\gamma^{2}}{3\beta}.
\end{equation}
From the equation (\ref{e4.9}) we have either $\gamma=0$ or $F_{S}=\frac{\alpha-\beta}{3}$. Now, if $\gamma=0$ then since $\gamma=\delta$, thus $\gamma=\delta=0$, implying the manifold reduces to a quasi-Einstein spacetime. Again if $F_{S}=\frac{\alpha-\beta}{3}$ then from the equation (\ref{e4.10}) we get $\gamma=0$ and hence $\gamma=\delta=0$, which again implies the manifold is reduced to a quasi-Einstein spacetime. This allows us to deduce the following theorem:\\\\
\textbf{Theorem 4.1:} {\em A Ricci pseudosymmetric $(HGQE)_{4}$
spacetime is a $(QE)_{4}$ spacetime if $\gamma=\delta$.}

\section{Projectively pseudosymmetric $(HGQE)_{4}$ spacetime} \label{s5}
From the equation $(\ref{e1.11})$ we see that for an $(HGQE)_{4}$
quasi-Einstein spacetime the $(0,4)$ projective curvature tensor
takes the following form,
\begin{equation} \label{e5.1}
P(X,Y,Z,W)=R(X,Y,Z,W)-\frac{1}{3}[S(Y,Z)g(X,W)-S(X,Z)g(Y,W)].
\end{equation}
A semi-Riemannian manifold $M^{n}(n \ge 3)$ is called projectively
pseudosymmetric if the following relation
\begin{equation}\label{e5.2}
(P(X,Y) \cdot S)(Z,W)=F_{S}Q(g,S)
\end{equation}
holds on $U_{S}=\{x \in M: S \neq \frac{r}{n}~at~x\}$ and $L_{S}$
is a function on $U_{S}$. From the equation (\ref{e5.2}) we get,
\begin{eqnarray}\label{e5.3}
&&~~S(P(X,Y)Z,W)+S(Z,P(X,Y)W)\nonumber\\
&&=F_{S}[g(Y,Z)S(X,W)-g(X,Z)S(Y,W)\nonumber\\
&&+g(Y,W)S(Z,X)-g(X,W)S(Y,Z)].
\end{eqnarray}
From the equation (\ref{e1.5}) from the equation (\ref{e5.3}) we
get,
\begin{eqnarray} \label{e5.4}
&&\alpha g(P(X,Y)Z,W)+\beta A(P(X,Y)Z)A(W)\nonumber\\
&&+\gamma[A(P(X,Y)Z)B(W)+B(P(X,Y)Z)A(W)] \nonumber\\
&&+\delta[A(P(X,Y)Z)D(W)+D(P(X,Y)Z)A(W)]\nonumber\\
&&+\alpha g(P(X,Y)W,Z)+\beta A(P(X,Y)W)A(Z)\nonumber\\
&&+\gamma [A(P(X,Y)W)B(Z)+B(P(X,Y)W)A(Z)] \nonumber\\
&&+\delta [A(P(X,Y)W)D(Z)+D(P(X,Y)W)A(Z)] \nonumber\\
&&=F_{S}[g(Y,Z)S(X,W)-g(X,Z)S(Y,W) \nonumber\\
&&~~+g(Y,W)S(Z,X)-g(X,W)S(Y,Z)].
\end{eqnarray}
Putting $X=\xi_{1}, Z=\xi_{2}, W=\xi_{3}$ in the equation
(\ref{e5.4}) we get,
\begin{eqnarray}\label{e5.5}
&&-\delta R(\xi_{1},Y,\xi_{2},\xi_{1})-\gamma R(\xi_{1},Y,\xi_{3},\xi_{1})\nonumber\\
&&=-\frac{\alpha \gamma}{3}D(Y)-\frac{\alpha \delta}{3}B(Y)+\frac{\gamma}{3}(\alpha D(Y)+\delta A(Y))\nonumber\\
&&+\frac{\delta}{3}(\alpha B(Y)+\gamma A(Y))-\frac{\gamma^{2}}{3}A(Y)-\frac{\delta^{2}}{3}A(Y)\nonumber\\
&&+F_{S}\{\gamma D(Y)+\delta B(Y)\}.
\end{eqnarray}
Contracting equation (\ref{e5.4}) over $X, W$ and putting
$Z=\xi_{1}$ we get,
\begin{eqnarray}\label{e5.6}
&&-\frac{\beta \gamma}{3}B(Y)-\frac{\beta \delta}{3}D(Y)-\frac{\gamma^{2}}{3}A(Y)\nonumber\\
&&+\frac{4\delta}{3}[\alpha D(Y)+\delta A(Y)]+\gamma(\beta-\alpha)B(Y)-\frac{\delta^{2}}{3}A(Y)\nonumber\\
&&-\frac{\delta r}{3}D(Y)+\delta(\beta-\alpha)D(Y)\nonumber\\
&&-\frac{4\alpha}{3}[(\alpha-\beta)A(Y)-\gamma B(Y)-\delta D(Y)]+\frac{\alpha r}{3}A(Y)\nonumber\\
&&+\frac{4\beta}{3}[(\alpha-\beta)A(Y)-\gamma B(Y)-\delta D(Y)]-\frac{\beta r}{3}A(Y)\nonumber\\
&&+\frac{4 \gamma}{3}[\alpha B(Y)+\gamma A(Y)]-\frac{\gamma r}{3}B(Y)\nonumber\\
&&-F_{S}\{rA(Y)-4[(\alpha-\beta)A(Y)-\gamma B(Y)-\delta D(Y)]\}\nonumber\\
&&=-\gamma R(\xi_{1},Y,\xi_{1},\xi_{2})-\delta
R(\xi_{1},Y,\xi_{1},\xi_{3}).
\end{eqnarray}
If $\gamma=\delta$ then rom the equations $(\ref{e5.5}),
(\ref{e5.6})$, by using the property of $R$ we get,
\begin{eqnarray}\label{e5.7}
&&-\frac{\beta \gamma}{3}B(Y)-\frac{\beta \gamma}{3}D(Y)-\frac{\gamma^{2}}{3}A(Y)\nonumber\\
&&+\frac{4\gamma}{3}[\alpha D(Y)+\gamma A(Y)]+\gamma(\beta-\alpha)B(Y)-\frac{\gamma^{2}}{3}A(Y)\nonumber\\
&&-\frac{\gamma r}{3}D(Y)+\gamma(\beta-\alpha)D(Y)\nonumber\\
&&-\frac{4\alpha}{3}[(\alpha-\beta)A(Y)-\gamma B(Y)-\gamma D(Y)]+\frac{\alpha r}{3}A(Y)\nonumber\\
&&+\frac{4\beta}{3}[(\alpha-\beta)A(Y)-\gamma B(Y)-\gamma D(Y)]-\frac{\beta r}{3}A(Y)\nonumber\\
&&+\frac{4 \gamma}{3}[\alpha B(Y)+\gamma A(Y)]-\frac{\gamma r}{3}B(Y)\nonumber\\
&&-F_{S}\{rA(Y)-4[(\alpha-\beta)A(Y)-\gamma B(Y)-\gamma D(Y)]\}\nonumber\\
&&=-\{-\frac{\alpha \gamma}{3}D(Y)-\frac{\alpha \gamma}{3}B(Y)+\frac{\gamma}{3}(\alpha D(Y)+\gamma A(Y))\nonumber\\
&&+\frac{\gamma}{3}(\alpha B(Y)+\gamma A(Y))-\frac{\gamma^{2}}{3}A(Y)-\frac{\gamma^{2}}{3}A(Y)\nonumber\\
&&+F_{S}\{\gamma D(Y)+\gamma B(Y)\}\}.
\end{eqnarray}
Putting $Y=\xi_{2}$ in equation (\ref{e5.7}) we get,
\begin{equation}\label{e5.8}
\frac{\gamma}{3}(-\beta+\alpha+3F_{S})=0.
\end{equation}
Again putting $Y=\xi_{1}$ in equation (\ref{e5.7}) we get,
\begin{equation}\label{e5.9}
3\beta F_{S}=\frac{2\gamma^{2}}{3}+\alpha \beta -\beta^{2}.
\end{equation}
From the equation (\ref{e5.8}) we have either $\gamma=0$ or $F_{S}=\frac{\alpha-\beta}{3}$. Now, if $\gamma=0$ then since $\gamma=\delta$, thus $\gamma=\delta=0$, implying the manifold reduces to a quasi-Einstein spacetime. Again if $F_{S}=\frac{\alpha-\beta}{3}$ then from the equation (\ref{e5.9}) we get $\gamma=0$ and hence $\gamma=\delta=0$, which again implies the manifold is reduced to a quasi-Einstein spacetime. This allows us to deduce the following theorem:\\\\
\textbf{Theorem 5.1:} {\em A projectively pseudosymmetric $(HGQE)_{4}$ spacetime is a $(QE)_{4}$ spacetime if $\gamma=\delta$.}\\\\

\section{Ricci symmetric $(HGQE)_{4}$ spacetime} \label{s6}
A semi-Riemannian manifold $M^{n}(n \ge 3)$ is called
Ricci-symmetric if $\nabla S=0$, where $S$ is the Ricci tensor of
the manifold. Considering the spacetime as a $(HGQE)_{4}$
spacetime we observe from equation $(\ref{e1.5})$ that the
manifold becomes Ricci symmetric if it satisfies the following
relation:
\begin{eqnarray}\label{e6.1}
&&\nabla _{Z}S(X,Y)=d\alpha(Z)g(X,Y)+d\beta(Z)A(X)A(Y)\nonumber\\
&&+\beta[(\nabla_{Z}A)(X)A(Y)+A(X)(\nabla_{Z}A)(Y)]\nonumber\\
&&+d\gamma(Z)[A(X)B(Y)+A(Y)B(X)]\nonumber\\
&&+\gamma[(\nabla_{Z}A)(X)B(Y)+A(X)(\nabla_{Z}B)(Y)]\nonumber\\
&&+(\nabla_{Z}A)(Y)B(X)+A(Y)(\nabla_{Z}B)(X)]\nonumber\\
&&+d\delta(Z)[A(X)D(Y)+A(Y)D(X)]\nonumber\\
&&+\delta[(\nabla_{Z}A)(X)D(Y)+A(X)(\nabla_{Z}D)(Y)]\nonumber\\
&&+(\nabla_{Z}A)(Y)D(X)+A(Y)(\nabla_{Z}D)(X)]=0.
\end{eqnarray}
Putting $X=Y=\xi_{1}$ in $(\ref{e6.1})$ we get,
\begin{equation}\label{6.2}
-d\alpha(Z)+d\beta(Z)-2\gamma(\nabla_{Z}B)(\xi_{1})-2\delta(\nabla_{Z}D)(\xi_{1})=0.
\end{equation}
Putting $X=Y=\xi_{2}$ in $(\ref{e6.1})$ we get,
\begin{equation}\label{6.3}
d\alpha(Z)+2\gamma(\nabla_{Z}A)(\xi_{2})=0.
\end{equation}
Again putting $X=Y=\xi_{3}$ in $(\ref{e6.1})$ we get,
\begin{equation}\label{6.4}
d\alpha(Z)+2\delta(\nabla_{Z}A)(\xi_{3})=0.
\end{equation}
Since the vector fields $\xi_{1}, \xi_{2}, \xi_{3}$ are mutually
orthogonal then $g(\xi_{1}, \xi_{2})= g(\xi_{1}, \xi_{3})= 0$,
this implies that $Z(g(\xi_{1}, \xi_{2}))= Z(g(\xi_{1},
\xi_{3}))=0$. Which further implies,
\begin{equation}\label{6.5}
(\nabla_{Z}B)(\xi_{1})=-(\nabla_{Z}A)(\xi_{2})
\end{equation}
and
\begin{equation}\label{6.6}
(\nabla_{Z}D)(\xi_{1})=-(\nabla_{Z}A)(\xi_{3}).
\end{equation}
Subtracting equations (\ref{6.3}),(\ref{6.4}) from (\ref{6.2}) and
using the relations (\ref{6.5}) and (\ref{6.6}) we get,
\begin{equation}\label{6.7}
d(\beta - 3\alpha)(Z)=0, ~\forall ~Z \in \rchi(M).
\end{equation}
That implies $\beta-3\alpha$ is a constant.\\\\
Now contracting the equation $(\ref{e6.1})$ over $X, W$ and using
$(\ref{6.5}), (\ref{6.6})$ we get,
\begin{equation}\label{6.8}
d(\beta - 4\alpha)(Z)=0, ~\forall ~Z \in \rchi(M).
\end{equation}
From the equations (\ref{6.7}) and (\ref{6.8}) it is clear that
$\alpha, \beta$ are constants. So, from the equations (\ref{6.3})
and (\ref{6.4}) we get
\begin{equation}\label{6.9}
\gamma(\nabla_{Z})(\xi_{1})=0
\end{equation}
and
\begin{equation}\label{6.10}
\delta(\nabla_{Z})(\xi_{2})=0.
\end{equation}
The equation (\ref{6.9}) shows $\gamma=0$ or
$\gamma(\nabla_{Z})(\xi_{1})=0$. If $\gamma=\delta$ then from
(\ref{e6.1}) we get,
\begin{equation}\label{6.11}
\beta[(\nabla_{Z}A)(X)A(Y)+A(X)(\nabla_{Z}A)(Y)=0.
\end{equation}
Putting $x=\xi_{1}$ in (\ref{6.11}) we get,
\begin{equation}\label{6.12}
\beta(\nabla_{Z}A)(Y)=0.
\end{equation}
If $\beta=0$ then since $\gamma=\delta=0$ the manifold reduces to
an Einstein manifold which is a contradiction. So, $\beta \neq 0$.
This implies that,
\begin{equation}\label{6.13}
(\nabla_{Z}A)(Y)=0.
\end{equation}
Again if $\gamma \neq 0$ then from (\ref{6.9}) we get,
\begin{equation}\label{6.14}
\beta(\nabla_{Z}A)(\xi_{2})=0.
\end{equation}
Using (\ref{6.14}) and putting $X=\xi_{1}, Y=\xi_{2}$ in the
equation (\ref{e6.1}) we get,
\begin{equation}\label{6.15}
d\gamma(Z)=0.
\end{equation}
Which imply $\gamma$ is also a constant. Then, putting $X=\xi_{2}$
in (\ref{e6.1}) and using (\ref{6.15}) we get,
\begin{equation}\label{6.16}
\gamma(\nabla_{Z}A)(Y)=0.
\end{equation}
Since $\gamma \neq 0$ thus we get the equation (\ref{6.13}) once
again. Hence, we always obtain $(\nabla_{Z}A)(Y)=0~\forall~Z,Y \in
\rchi(M).$ Which can be written as
 \begin{equation}\label{6.17}
g(\nabla_{Z}\xi_{1},Y)=0~\forall~Z,Y \in \rchi(M).
\end{equation}
Thus we obtain
\begin{equation}\label{6.18}
\nabla_{Z}\xi_{1}=0.
\end{equation}
Which implies that the generator vector field $\xi_{1}$ is always parallel. Hence we obtain the following theorem as:\\\\
\textbf{Theorem 6.1:} {\em In a Ricci symmetric $(HGQE)_{4}$ spacetime with $\gamma=\delta$ the generator vector field $\xi_{1}$ is always parallel}.\\\\
Again putting $Z=\xi_{1}$ in (\ref{6.18}) we get,
\begin{equation}\label{6.19}
\nabla_{\xi_{1}}\xi_{1}=0.
\end{equation}
Which implies that the integral curves of $\xi{1}$ are geodesics. This leads us to the next theorem as:\\\\
\textbf{Theorem 6.2:} {\em In a Ricci symmetric $(HGQE)_{4}$ spacetime with $\gamma=\delta$ the integral curves of the generator vector field $\xi_{1}$ are geodesics}.\\\\
With the help of the theorems (6.1) and (6.2) we arrive at the
following condition
\begin{equation}\label{6.20}
R(X,Y)\xi_{1}=\nabla_{X}\nabla_{Y}\xi_{1}-\nabla_{Y}\nabla_{X}\xi_{1}-\nabla_{[X,Y]}\xi_{1}=0.
\end{equation}
Hence we obtain the following theorem as:\\\\
\textbf{Theorem 6.3:} {\em In a Ricci symmetric $(HGQE)_{4}$ spacetime with $\gamma=\delta$ the Riemannian curvature tensor vanishes at the generator vector field}.\\\\
Now contracting (\ref{6.20}) we get
\begin{equation}\label{6.21}
S(X,\xi_{1})=0.
\end{equation}
Thus from (\ref{e1.5}) we get,
\begin{equation}\label{6.22}
(\alpha-\beta)A(X)-\gamma[B(X)-D(X)]=0.
\end{equation}
Since $\gamma=\delta$ thus putting $X=\xi_{1}$ in (\ref{6.22}) we
get,
\begin{equation}\label{6.23}
\alpha=\beta.
\end{equation}
Again putting $X=\xi_{2}$ in (\ref{6.22}) we get,
\begin{equation}\label{6.24}
\gamma=0.
\end{equation}
Thus $\gamma=\delta=0$, hence from (\ref{e1.5}) we get,
\begin{equation}\label{6.25}
S(X,Y)=\alpha[g(X,Y)+A(X)A(Y)].
\end{equation}
This allows us to arrive at the next theorem as:\\\\
\textbf{Theorem 6.4:} {\em Every Ricci symmetric $(HGQE)_{4}$ spacetime with $\gamma=\delta$ is a $(QE)_{4}$ spacetime with the scalar functions are constants and equal}.\\\\
Using the equations (\ref{6.25}) from the equation (\ref{e1.9}) we
get,
\begin{equation}\label{6.26}
T(X,Y)=\frac{2\lambda-\alpha}{2k} g(X,Y)+\frac{\alpha}{k}A(X)A(Y).
\end{equation}
Since $\alpha, \lambda, k$ all are constants thus taking
derivative to both the sides the equation (\ref{6.26}) we get,
\begin{equation}\label{6.27}
(\nabla_{Z}T)(X,Y)=0.
\end{equation}
Therefore we see in this case the energy-momentum tensor is covariantly constant. This leads us to the following theorem:\\\\
\textbf{Theorem 6.5:} {\em In a Ricci-symmetric $(HGQE)_{4}$ spacetime with $\gamma=\delta$ satisfying Einstein field equation with cosmological constant the energy-momentum tensor is covariantly constant.}\\\\
Again from (\ref{6.26}) we observe that since the velocity vector field $\xi_{1}$ is parallel and $\alpha$ is a constant thus the energy-momentum tensor is of Codazzi type. Hence we derive the next theorem as:\\\\
\textbf{Theorem 6.6:} {\em In a Ricci-symmetric $(HGQE)_{4}$ spacetime with $\gamma=\delta$ satisfying Einstein field equations with cosmological constant the energy-momentum tensor is of Codazzi type}.\\\\
Now from the equations (\ref{e1.8}), (\ref{e1.9}) and theorem
(6.4) we conclude the values of $\sigma$ and $p$ as,
\begin{equation}\label{6.28}
\sigma=\frac{3\alpha-2\lambda}{2k},~p=\frac{2\lambda-\alpha}{2k}.
\end{equation}
Since, $\alpha, \lambda, k$ all are constants thus we get $\sigma, p$ are also constants. This leads us to the following theorem:\\\\
\textbf{Theorem 6.7:} {\em In a Ricci-symmetric $(HGQE)_{4}$ spacetime with $\gamma=\delta$ satisfying Einstein field equation with cosmological constant the energy density and the isotropic pressure are constants}.\\\\
It is proved $\cite{jmh4}$ that in a perfect fluid spacetime if the energy-momentum tensor is of Codazzi type then the vorticity and shear of the spacetime vanish. Hence we derive the next theorem:\\\\
\textbf{Theorem 6.8:} {\em In a Ricci-symmetric $(HGQE)_{4}$ spacetime with $\gamma=\delta$ satisfying Einstein field equations with cosmological constant the vorticity and the shear tensor vanish.}\\\\
Here we see that the velocity vector $\xi_{1}$ is constant over the spacelike hypersurface orthogonal to $\xi_{1}$. But it is described in $\cite{jmh23}$ that perfect fluid spacetime that is vorticity free and shear free is of petrov type $I, D$ or $O$. Thus we state the next theorem as:\\\\
\textbf{Theorem 6.9:} {\em The local cosmological structure of a Ricci-symmetric $(HGQE)_{4}$ spacetime with $\gamma=\delta$ satisfying Einstein field equation with cosmological constant can be identified as petrov type $I, D$ or $O$}.\\\\

\section{$(HGQE)_{4}$ spacetime with cyclic parallel Ricci tensor} \label{s7}
Consider an $(HGQE)_{4}$ with cyclic parallel Ricci tensor. Then
we get the following equation,
\begin{equation}\label{e7.1}
(\nabla_{X}S)(Y,Z)+(\nabla_{Y}S)(X,Z)+(\nabla_{Z}S)(X,Y)=0.
\end{equation}
From the equation $(\ref{e1.9})$ we have
\begin{equation}\label{e7.2}
(\nabla_{X}S)(Y,Z)=\frac{1}{2}dr(Z)g(X,Y)+k(\nabla_{Z}T)(X,Y).
\end{equation}
Now, if in an $(HGQE)_{4}$ with cyclic parallel Ricci tensor the
scalar curvature of the spacetime is constant then,
\begin{equation}\label{e7.3}
dr(X)=0,
\end{equation}
for all $X \in \rchi(M)$. Using the equation $(\ref{e7.3})$ in the
equation $(\ref{e8.2})$ we get,
\begin{equation}\label{e7.4}
(\nabla_{X}S)(Y,Z)=k(\nabla_{Z}T)(X,Y).
\end{equation}
Now, if the $(HGQE)_{4}$ spacetime is cyclic parallel then,
\begin{eqnarray}\label{e7.5}
&&k\{(\nabla_{X}T)(Y,Z)+(\nabla_{Y}T)(X,Z)+(\nabla_{Z}T)(X,Y)\}\nonumber\\
&&=(\nabla_{X}S)(Y,Z)+(\nabla_{Y}S)(X,Z)+(\nabla_{Z}S)(X,Y)=0.
\end{eqnarray}
Since $k$, being the gravitational constant is always nonzero,
from the equation (\ref{e8.5}) we have
\begin{equation}\label{e7.6}
(\nabla_{X}T)(Y,Z)+(\nabla_{Y}T)(X,Z)+(\nabla_{Z}T)(X,Y)=0.
\end{equation}
This allows us to obtain the following theorem as:\\\\
\textbf{Theorem 7.1:} {\em In a hyper-generalised quasi-Einstein
spacetime with cyclic parallel Ricci tensor if the scalar
curvature is constant then the energy-momentum tensor is also
cyclic parallel.}

\section{On the physical applications of an $(HGQE)_{4}$ spacetime} \label{s8}
Here we study some physical applications of the $(HGQE)_{4}$
spacetime. In $\cite{jmh33}$, $\cite{jmh12}$ G. F. R. Ellis has
given the energy momentum tensor of a fluid matter distribution as
follows:
\begin{eqnarray}\label{e8.1}
&&T(X,Y)=(\sigma+p)A(X)A(Y)+pg(X,Y)+A(X)B(Y)+A(Y)B(X)\nonumber\\
&&~~~~~~~~~~~~~~+A(X)D(Y)+A(Y)D(X),
\end{eqnarray}
where,\\
$~~~~~~~~~~~~~g(X,\xi_{1})=A(X),~g(X,\xi_{2})=B(X),~g(X,\xi_{3})=D(X),$\\
$~~~~~~~~~~~~~A(\xi_{1})=-1,~B(\xi_{2})=1,~D(\xi_{3})=1,$\\
$~~~~~~~~~~~~~g(\xi_{1},~\xi_{2})=0,~g(\xi_{2},~\xi_{3})=0,~g(\xi_{3},~\xi_{1})=0,$\\
and $\sigma$ is the matter density, $p$ is the isotropic pressure, $\xi_{1}$ is the timelike velocity vector field, $\xi_{2}$ is the heat conduction vector field and $\xi_{3}$ is the stress vector field.\\\\
Combining equation (\ref{e8.1}) with equation (\ref{e1.9}) we get,
\begin{eqnarray}\label{e8.2}
&&S(X,Y)=(kp+\frac{r}{2}-\lambda)g(X,Y)+k(\sigma+p)A(X)A(Y)\nonumber\\
&&~~~~~~~~~~~~~~+k[A(X)B(Y)+A(Y)B(X)]\nonumber\\
&&~~~~~~~~~~~~~~+k[A(X)D(Y)+A(Y)D(X)].
\end{eqnarray}
Comparing equation (\ref{e8.2}) with equation (\ref{e1.9}) we get that this spacetime is clearly an $(HGQE)_{4}$ spacetime with the constants $\gamma=\delta=k.$ Hence all the results that we derived in the earlier sections are absolutely effective in this spacetime, and thus we derive the following set of theorems:\\\\
From theorem (2.1) we get,\\\\
\textbf{Theorem 8.1:} {\em A $W_{2}$-Ricci pseudosymmetric viscous fluid $(HGQE)_{4}$ spacetime is a $(QE)_{4}$ spacetime.}\\\\
From theorem (3.1) we get,\\\\
\textbf{Theorem 8.2:} {\em A $Z$-Ricci pseudosymmetric viscous fluid $(HGQE)_{4}$ spacetime is a $(GQE)_{4}$ spacetime.}\\\\
From theorem (4.1) we get,\\\\
\textbf{Theorem 8.3:} {\em A Ricci pseudosymmetric viscous fluid $(HGQE)_{4}$ spacetime is a $(QE)_{4}$ spacetime.}\\\\
From theorem (5.1) we get,\\\\
\textbf{Theorem 8.4:} {\em A projectively pseudosymmetric viscous fluid $(HGQE)_{4}$ spacetime is a $(QE)_{4}$ spacetime.}\\\\
From theorem (6.1) we get,\\\\
\textbf{Theorem 8.5:} {\em In a Ricci symmetric $(HGQE)_{4}$ viscous fluid spacetime the generator vector field $\xi_{1}$ is always parallel}.\\\\
From theorem (6.2) we get,\\\\
\textbf{Theorem 8.6:} {\em In a Ricci symmetric $(HGQE)_{4}$ viscous fluid spacetime the integral curves of the generator vector field $\xi_{1}$ are geodesics}.\\\\
From theorem (6.3) we get,\\\\
\textbf{Theorem 8.7:} {\em In a Ricci symmetric $(HGQE)_{4}$ viscous fluid spacetime the Riemannian curvature tensor vanishes at the generator vector field}.\\\\
From theorem (6.4) we get,\\\\
\textbf{Theorem 8.8:} {\em Every Ricci symmetric $(HGQE)_{4}$ viscous fluid spacetime is a $(QE)_{4}$ spacetime with the scalar functions are constants and equal}.\\\\
From theorem (6.5) we get,\\\\
\textbf{Theorem 8.9:} {\em In a Ricci symmetric $(HGQE)_{4}$ viscous fluid spacetime satisfying Einstein field equation with cosmological constant the energy-momentum tensor is covariantly constant}.\\\\
From theorem (6.6) we get,\\\\
\textbf{Theorem 8.10:} {\em In a Ricci symmetric $(HGQE)_{4}$ viscous fluid spacetime satisfying Einstein field equation with cosmological constant the energy-momentum tensor is of Codazzi type}.\\\\
From theorem (6.7) we get,\\\\
\textbf{Theorem 8.11:} {\em In a Ricci symmetric $(HGQE)_{4}$ viscous fluid spacetime satisfying Einstein field equation with cosmological constant the energy density and the isotropic pressure are constants}.\\\\
From theorem (6.8) we get,\\\\
\textbf{Theorem 8.12:} {\em In a Ricci symmetric $(HGQE)_{4}$ viscous fluid spacetime satisfying Einstein field equation with cosmological constant the vorticity and the shear tensor vanish}.\\\\
From theorem (6.9) we get,\\\\
\textbf{Theorem 8.13:} {\em The local cosmological structure of a Ricci-symmetric $(HGQE)_{4}$ viscous fluid spacetime satisfying Einstein field equation with cosmological constant can be identified as petrov type $I, D$ or $O$}.\\\\

\section{Example of $(HGQE)_{4}$ spacetime} \label{s9}
Finally we give a non-trivial example to establish the existence of $(HGQE)_{4}$ spacetime non-trivially. For this we consider a metric known as Lorentzian metric $g$ on $M^4$ given by \\\\
$ds^2 = g_{ij}dx^i dx^j = -\frac{k}{r}(dt)^2+ \frac{1}{\frac{c}{r}-4}(dr)^2+ r^2 (d\theta)^2 + (r \sin\theta)^2(d\phi)^2,$\\\\
where $i,j = 1,2,3,4$ and $k,c$ are constants. Thus we obtain the  nonzero components of Christofell symbols, curvature tensors and Ricci tensors as follows:\\\\
\begin{eqnarray}\label{9.1}
\Gamma_{33}^2=4r-c, \Gamma_{12}^1=-\frac{1}{2r}, \Gamma_{22}^2=\frac{c}{2r(c-4r)}, \Gamma_{32}^3=\Gamma_{42}^4=\frac{1}{r},\nonumber\\
\Gamma_{43}^4=\cot\theta, \Gamma_{44}^2=(4r-c)(\sin\theta)^2,
\Gamma_{44}^3=-\frac{\sin(2\theta)}{2}
\end{eqnarray}
\begin{eqnarray}\label{9.2}
R_{1221}=-\frac{k(c-3r)}{r^{3} (c-4r)},  R_{1331}=\frac{k(c-4r)}{2r^{2}}, R_{1441}=\frac{k(c-4r)(\sin\theta)^{2}}{2r^{2}}\nonumber\\
R_{2332}= \frac{c}{2(4r-c)}, R_{2442}=\frac{c(\sin\theta)^{2}}{2(4r-c)}, R_{3443}= r(c-5r)(\sin\theta)^{2}\nonumber\\
R_{11}= -\frac{k}{r^{3}}, R_{22}= -\frac{3}{r(c-4r)}, R_{33}= -3,
R_{44}= -3(\sin\theta)^{2}
\end{eqnarray}
From $(\ref{9.1})$ and $(\ref{9.2})$ it follows that $M^4$ is a Lorentzian manifold of nonzero scalar curvature ($= -\frac{8}{r^2}$). Now we will prove that this is an $(HGQE)_4$ manifold.\\\\
We consider $\alpha, \beta, \gamma$ and $\delta$ as the associated
scalars and we consider them as follows:
\begin{equation}
\alpha=-\frac{5}{r^{2}}, \beta=-\frac{12}{r^{2}},
\gamma=\frac{3}{r^{2}}, \delta=-\frac{4}{r^{2}}
\end{equation}
and the associated $1$-forms are as follows :\\\\
\begin{math}
~~~~~~~~~~~~A_{i}(x)=\left \{\begin{array}{cccl}
\sqrt{\frac{k}{2r}} & \mbox{for} & i=1 \\
\sqrt{\frac{r}{6(c-4r)}} & \mbox{for} & i=2\\
0 & \mbox{for} & i=3,4\\
\end{array}
\right.
\end{math}~~~~~~;~~~~~~~\begin{math}B_{i}(x)=\left
\{\begin{array}{ccl}
\sqrt{\frac{k}{r}} & \mbox{for} & i=1 \\
0 & \mbox{for} & i= 2,3,4\end{array} \right.
\end{math}\\
\begin{math}~~~~~~~~~~D_{i}(x)=\left
\{\begin{array}{ccl}
\sqrt{\frac{k}{r}} & \mbox{for} & i=1 \\
0 & \mbox{for} & i= 2,3,4\end{array} \right.
\end{math}\\\\\\

Hence we gain,\\\\
$(i) R_{11}= \alpha g_{11}+\beta A_{1}A_{1}+\gamma[A_{1}B_{1}+B_{1}A_{1}]+\delta[A_{1}D_{1}+D_{1}A_{1}]$ \\\\
$(ii) R_{22}= \alpha g_{22}+\beta A_{2}A_{2}+\gamma[A_{2}B_{2}+B_{2}A_{2}]+\delta[A_{2}D_{2}+D_{2}A_{2}]$ \\\\
$(iii) R_{33}= \alpha g_{33}+\beta A_{3}A_{3}+\gamma[A_{3}B_{3}+B_{3}A_{3}]+\delta[A_{3}D_{3}+D_{3}A_{3}]$ \\\\
$(iv) R_{44}= \alpha g_{44}+\beta A_{4}A_{4}+\gamma[A_{4}B_{4}+B_{4}A_{4}]+\delta[A_{4}D_{4}+D_{4}A_{4}]$ \\\\
As every Ricci tensor other than $R_{11}, R_{22}, R_{33}$ and $R_{44}$ are zero, so we obtain \\\\
~~~$R_{ij}= \alpha g_{ij}+\beta A_{i}A_{j}+\gamma[A_{i}B_{j}+B_{i}A_{j}]+\delta[A_{i}D_{j}+D_{i}A_{j}], i,j= 1,2,3,4.$\\\\
Consequently, scalar curvature $=4\alpha-\beta=-\frac{8}{r^{2}}$. Hence, $(M^{4}, g)$ is a hyper-generalized quasi Einstein manifold. \\\\
\textbf{Conclusion:} {\em The general theory of relativity is the most prominent flagship of modern physics. It deals with the curvature of spacetime. As  hyper-generalized quasi-Einstein spacetime is considered as the base space of the fluid matter distribution. Thus it has been very necessary to study about the geometric and physical applications of hyper-generalized quasi-Einstein spacetime. It deals with the relativistic viscous fluid spacetime admitting heat flux and stress. The general theory of relativity describes gravity as a geometric property of spacetime. The curvature of spacetime is directly related to the energy-momentum tensor. Also we know the cosmological constant to be of homogeneous energy density which causes the expansion of the universe to accelerate. So, here we obtain a set of geometric and physical properties of hyper-generalized quasi-Einstein spacetimes and give a non-trivial example to establish the existence of hyper-generalized quasi-Einstein spacetime non-trivially.}\\\\

\end{document}